\documentclass[prd,twocolumn,showpacs,showkeys,footinbib]{revtex4}
\usepackage{amsmath}
\usepackage{graphicx}
\usepackage{slashed}

\renewcommand\vec[1]{{\boldsymbol{\bf #1}}}
\newcommand\adj[1]{\bar{#1}}
\newcommand\he[1]{#1^{\dagger}}
\DeclareMathOperator{\im}{Im}
\DeclareMathOperator{\tr}{Tr}
\DeclareMathOperator{\sgn}{sgn}
\newcommand\La{\mathcal L}
\newcommand\D{\mathcal D}
\newcommand\eps{\epsilon}
\newcommand\cc[1]{{#1}^{\mathcal C}}

\begin{document}

\title{BCS--BEC crossover in dense relativistic matter: Collective excitations}
\author{Tom\'{a}\v{s} Brauner}
\email{brauner@ujf.cas.cz}
\affiliation{Institut f\"ur Theoretische Physik, J. W. Goethe-Universit\"at,
Max von Laue-Str. 1, 60438 Frankfurt am Main, Germany\footnote{On
leave from Department of Theoretical Physics, Nuclear Physics Institute ASCR,
25068 \v Re\v z, Czech Republic}}

\begin{abstract}
We study the relativistic BCS--BEC crossover within a class of
Nambu--Jona-Lasinio type models, including arbitrary Lorentz-scalar pairing
channels. Using the mean-field approximation we investigate spectral properties
of the collective bosonic excitations in the superfluid phase, with particular
attention to the Nambu--Goldstone bosons of the broken symmetry. This is a
first step towards a systematic improvement of the mean-field approximation by
including the fluctuation effects. The general results are illustrated on pairing in
dense two-flavor quark matter --- the two-flavor color superconductor.
\end{abstract}

\maketitle

\section{Introduction}
It was noted long ago that Bardeen--Cooper--Schrieffer (BCS) superconductivity
of correlated Cooper pairs and Bose--Einstein condensation (BEC) of tightly
bound difermion molecules are merely two sides of the same coin. Eagles
\cite{Eagles:1969ea} and Leggett \cite{Leggett:1980le} were perhaps the first to observe
that a many-fermion system can display both the BCS and the BEC behavior, and these two
regimes are connected by a smooth crossover as the strength of the attractive interaction
between the fermions is varied. The crossover can be described at least qualitatively well
by the conventional mean-field approximation provided one determines the chemical
potential self-consistenly by fixing the overall particle density. 

However, it was only recently that the BCS--BEC crossover could be realized experimentally
using ultracold atomic Fermi gases
\cite{Regal:2004re,Bartenstein:2004ba,Zwierlein:2004zw,Bourdel:2004bo}, where the strength
of the interaction can be tuned by external magnetic field and thereby, using the Feshbach
resonance, cover the whole range between the BCS and BEC limits. The dilute atomic gases
represent particularly clean and well controlled examples of interacting many-body
systems. It is believed that they may provide an ideal theoretical playground for
development of many-body techniques, that would be later applied to more complicated
strongly correlated systems. 

The mean-field theory of Eagles and Leggett was for the first time improved by Nozi\`eres
and Schmitt-Rink \cite{Nozieres:1985no} who included the contribution of the pair
fluctuations to the total particle density. In the past decade, various extensions of the
mean-field approximation have been proposed, including boson--fermion models
\cite{Romans:2006ro} as well as purely fermionic self-consistent schemes
\cite{Haussmann:1994ha,Haussmann:2007ha}. An extensive list of references may
be found in the reviews \cite{Loktev:2001lo,Chen:2005ch}. The most recently developed
techniques are compared in \cite{Diener:2007di}. Simultaneously to the analytic
approximation schemes, the interacting Fermi gas is also being studied using numerical
Monte Carlo simulations \cite{Carlson:2003ca,Burovski:2006bu}. 

In high-energy physics, the analogy with BCS superconductivity was used long ago by
Nambu and Jona-Lasinio \cite{Nambu:1961tp,Nambu:1961fr} to propose a model for
dynamical chiral symmetry breaking. Later, Cooper pairing of two fermions near their Fermi
surface found an application in the physics of dense nuclear/quark matter, giving rise in
particular to the phenomenon of color superconductivity
\cite{Barrois:1977xd,Frautschi:1978rz}; see \cite{Alford:2007rw} for the most recent
review. The crossover physics is relevant for these high-energy systems because of the
strong-coupling nature of quantum chromodynamics. Concretely, the so-called pseudogap
phase as a precursor to superfluidity of tightly bound Cooper pairs was investigated in
models of chiral symmetry breaking
\cite{Babaev:1999iz,Babaev:2000fj,Hands:2001cs,Castorina:2005tm} as well as color
superconductivity \cite{Kitazawa:2003cs}. The structural change of Cooper pairs with
increasing coupling strength and the possibility of their BEC were studied in
\cite{Matsuzaki:1999ww,Abuki:2001be,Nawa:2005sb,Kitazawa:2007zs}. The spectrum of diquarks
in various color-superconducting phases was investigated by Ebert \emph{et al.}
\cite{Ebert:2004dr,Ebert:2007bp}.

The actual crossover between the BCS and BEC regimes in dense quark matter in dependence
on the coupling strength has started to be studied only recently
\cite{Deng:2006ed,He:2007yj,Sun:2007fc}. Perhaps the first attempts to go beyond the
mean-field approximation have been made in Refs.
\cite{Nishida:2005ds,Abuki:2006dv,He:2007kd}. In particular, Abuki \cite{Abuki:2006dv} has
conducted an extensive study of fluctuation effects in the normal phase, using the
Gaussian approximation. The goal of this paper is to investigate the spectrum of the
bosonic collective modes below the critical temperature. Thereby we construct a framework
for description of fluctuation effects in the superfluid phase.

The plan of the paper is following. In the next section, we derive the general
formulas valid for a class of models of the Nambu--Jona-Lasinio (NJL) type,
that describe (even-parity) spin-zero pairing in relativistic fermion matter. The model of
Abuki \cite{Abuki:2006dv}, which is a special case, is then investigated in
detail in the following section. In particular, we study the spectral
properties of the bosonic collective modes in the mean-field approximation.
Possible extensions of the present work are discussed in the conclusions.

\section{NJL models of pairing}
\subsection{Model definition}
We consider the general class of models defined by the Lagrangian
\begin{equation}
\La=\adj\psi(i\slashed\partial+\mu\gamma_0-m)\psi+\frac G4\sum_a
(\psi^TT_a\psi)(\adj\psi\adj T_a\adj\psi^T),
\label{Lagrangian}
\end{equation}
where $\adj T_a=\gamma_0\he T_a\gamma_0$. The spinor $\psi$ denotes a set of
fermion flavors with common mass $m$ and chemical potential $\mu$, while $T_a$
are matrices that furnish an antisymmetric tensor representation under the
symmetry transformations acting on $\psi$. In general, they act on both Dirac
and flavor indices of $\psi$. Even though the Lagrangian \eqref{Lagrangian}
covers, in a simplified manner, a number of systems investigated in literature
\cite{Buballa:2003qv}, at this stage we completely neglect correlations in the
fermion--antifermion channel that may lead to dynamical breaking of the chiral symmetry.

This model is analyzed using the standard Hubbard--Stratonovich transformation.
We start with the definition of the Nambu spinor $\Psi=(\psi,\cc\psi)^T$
\footnote{For sake of an easy comparison, we closely follow the notation of
Abuki \cite{Abuki:2006dv}. The only important difference is that we define the
Nambu doublet with the charge conjugate spinor $\cc\psi$ instead of
the simple Dirac conjugate $\adj\psi^T$.}. Using the charge conjugate field,
$\cc\psi=C\adj\psi^T$, the interaction term may be rewritten as $\frac
G4\left|\cc{\adj\psi}P_a\psi\right|^2$, where $P_a=C^{-1}T_a$. It is removed
from the Lagrangian by introducing new auxiliary scalar fields $\Delta_a$ and
adding
$$
\Delta\La=-\frac1G\left|\Delta_a-\frac G2\adj\Psi\left(\begin{matrix}
0 & 0\\
P_a & 0
\end{matrix}\right)\Psi\right|^2.
$$
The new, semibosonized Lagrangian reads
$$
\La_{\text{semi}}=-\frac1G\Delta_a^*\Delta_a+\frac12\adj\Psi\D^{-1}\Psi,
$$
where
\begin{equation}
\D^{-1}=\left(\begin{matrix}
i\slashed\partial+\gamma_0\mu-m & \Delta_a\adj P_a\\
\Delta^*_aP_a & i\slashed\partial-\gamma_0\mu-m
\end{matrix}\right),
\label{fermion_propagator_coord}
\end{equation}
is the fermion propagator in the coordinate space. Upon integrating out the
fermions, the thermodynamic potential $\Omega$ is given by
\begin{equation}
\begin{split}
e^{-\beta\Omega}&=\int
d\Delta_ad\Delta^*_a\exp\left(-S_{\text{eff}}[\Delta,\Delta^*]\right),\\
S_{\text{eff}}&=\frac1G\int_0^{\beta}d\tau\int d^3\vec
x\,|\Delta_a(\vec x,\tau)|^2-\frac12\log\det\D^{-1}.
\end{split}
\label{Action_bosonized}
\end{equation}

This functional integral, of course, cannot be evaluated exactly. Note that the
bosonized action $S_{\text{eff}}$ includes the effect of fermion loops to all
orders, due to the presence of the determinant of the inverse fermion
propagator $\D^{-1}$. In other words, it contains all Feynman graphs with
the scalar fields as external lines. It is thus to be interpreted as a
classical action for the scalars $\Delta_a,\Delta^*_a$, and the above
functional integral as the generating functional of their Green's functions
(at zero source). This can be determined from the quantum effective action
associated to the classical action $S_{\text{eff}}$, being the sum of all
one-boson-irreducible graphs.

Since we assume that the leading nonperturbative behavior, which gives rise to
the pairing, is borne by the fermion loops that have already been resummed to
all orders, the expansion in the number of \emph{bosonic} loops provides a
natural approximation scheme for the calculation of the thermodynamic
potential. The tree level corresponds to simply setting
$\beta\Omega_{\text{MF}}=S_{\text{eff}}$: This is the standard mean-field
approximation. At one loop, the thermodynamic potential would be given by
\begin{equation}
\beta\Omega_{\text{1L}}=\beta\Omega_{\text{MF}}+\frac12\log\det\Xi,
\label{one_loop_TDpot}
\end{equation}
where
\begin{equation}
\Xi_{ab}(X-Y)=\frac{\delta^2S_{\text{eff}}}{\delta\he\Phi_a(X)\delta\Phi_b(Y)}\equiv
\left(\begin{matrix}
\chi_{ab}^{\Delta\Delta^*} & \chi_{ab}^{\Delta\Delta}\\
\chi_{ab}^{\Delta^*\Delta^*} & \chi_{ab}^{\Delta^*\Delta}
\end{matrix}\right),
\label{Xi_matrix}
\end{equation}
is the inverse propagator of the bosonic modes; $\Phi$ is the scalar doublet
field, $\Phi_a=(\Delta_a,\Delta^*_a)^T$. Note that in the normal phase, Eq.
\eqref{one_loop_TDpot} reduces to the Gaussian approximation adopted by Abuki
\cite{Abuki:2006dv}. Carrying out the indicated differentiation and setting the
fields $\Delta_a$ equal to their vacuum expectation values (which we hereafter
refer to by the same symbol), we arrive at the expressions for the normal and
anomalous parts of the inverse propagators,
\begin{multline}
\chi_{ab}^{\Delta\Delta^*}(i\Omega_N,\vec
p)=\frac1G\delta_{ab}+\frac12T\sum_n\int\frac{d^3\vec
k}{(2\pi)^3}\\
\times\tr\left[P_a\D^{\psi\adj\psi}(i\omega_n,\vec k)\adj P_b
\D^{\adj\psi\psi}(i\omega_n-i\Omega_N,\vec k-\vec p)\right],
\label{chi_general_norm}
\end{multline}
\begin{multline}
\chi_{ab}^{\Delta\Delta}(i\Omega_N,\vec p)=\frac12T\sum_n\int\frac{d^3\vec
k}{(2\pi)^3}\\
\times\tr\left[P_a\D^{\psi\psi}(i\omega_n,\vec k)P_b
\D^{\psi\psi}(i\omega_n-i\Omega_N,\vec k-\vec p)\right],
\label{chi_general_anom}
\end{multline}
where $\psi$ and $\adj\psi$ in the superscripts to $\D$ denote the respective
matrix elements of the matrix propagator \eqref{fermion_propagator_coord}. The
remaining two entries of the matrix \eqref{Xi_matrix} follow from the general
relations,
\begin{align*}
\chi^{\Delta^*\Delta}_{ab}(\omega,\vec p)&=\chi^{\Delta\Delta^*}_{ba}(-\omega,-\vec p),\\
\chi^{\Delta\Delta}_{ab}(\omega,\vec p)&=\chi^{\Delta\Delta}_{ba}(-\omega,-\vec p),\\
\chi^{\Delta^*\Delta^*}_{ab}(\omega,\vec p)&=\left[\chi^{\Delta\Delta}_{ba}(\omega^*,\vec
p)\right]^*.
\end{align*}
valid for an arbitrary complex frequency $\omega$.

\subsection{Fermion propagator}
In order to be actually able to invert the matrix $\D^{-1}$
\eqref{fermion_propagator_coord}, we now make a specific assumption about the
fermion pairing pattern. We assume that the Cooper pairs are \emph{Lorentz
scalars}, i.e., the matrices $T_a$ have the form $T_a=C\gamma_5Q_a$, where
$Q_a$ act just on the internal symmetry indices. The order parameters
$\Delta_a$ now enter the fermion propagator in terms of the expression
$M=\Delta_a\he Q_a$. Note that Pauli principle requires $T_a$ to be
antisymmetric so that both $Q_a$ and $M$ have to be symmetric matrices.

With the above assumption, the denominators of the matrix fermion propagator
depend only on the combinations $M\he M$ and $\he MM$. Being Hermitian and
positive definite, the matrix $M\he M$ may be spectrally decomposed as
\cite{Rischke:2004rw}
$$
M\he M=\sum_r\Delta_r^2\mathcal P_r,
$$
where $\mathcal P_r$ is a set of projectors on the respective eigenvectors, and
$\Delta_r^2$ are the real positive eigenvalues whose relation to $\Delta_a$
will be clarified later. Using the symmetry of $M$, we find the equivalent
decomposition,
$$
\he MM=\sum_r\Delta_r^2\mathcal P_r^*.
$$

The parameters $\Delta_r$ play the role of gaps in the (fermionic)
quasiparticle dispersion relations, which are found as poles in the fermion
propagator $\D$,
\begin{equation}
\begin{split}
E^e_{\vec kr}&=\sqrt{(\xi^e_{\vec k})^2+\Delta_r^2},\\
\text{with}\quad\xi^e_{\vec k}&=\eps_{\vec k}+e\mu,\quad \eps_{\vec
k}=\sqrt{\vec k^2+m^2},\quad e=\pm.
\end{split}
\label{E_xi_definition}
\end{equation}
Explicit expressions for the matrix elements of the fermion propagator, obtained by
inverting Eq. \eqref{fermion_propagator_coord}, read,
\begin{equation}
\begin{split}
\D^{\psi\adj\psi}(i\omega_n,\vec k)
&=\sum_r\sum_{e=\pm}\frac{i\omega_n-e\xi^e_{\vec k}}{(i\omega_n)^2-(E^e_{\vec
k r})^2}\Lambda^{-e}_{\vec k}\gamma_0\mathcal P_r,\\
\D^{\adj\psi\psi}(i\omega_n,\vec k)
&=\sum_r\sum_{e=\pm}\frac{i\omega_n+e\xi^e_{\vec k}}{(i\omega_n)^2-(E^e_{\vec
k r})^2}\Lambda^{e}_{\vec k}\gamma_0\mathcal P_r^*,\\
\D^{\psi\psi}(i\omega_n,\vec k)
&=-\sum_r\sum_{e=\pm}\frac{\Lambda^{-e}_{\vec k}\gamma_5}
{(i\omega_n)^2-(E^e_{\vec k r})^2}\mathcal P_rM,\\
\D^{\adj\psi\adj\psi}(i\omega_n,\vec k)
&=\sum_r\sum_{e=\pm}\frac{\Lambda^{e}_{\vec k}\gamma_5}{(i\omega_n)^2-(E^e_{\vec
k r})^2}\he M\mathcal P_r.\\
\end{split}
\label{fermion_propagator}
\end{equation}
The definition and basic properties of the standard energy projectors $\Lambda_{\vec k}^e$
are given in Appendix \ref{App:projectors}.

\subsection{Propagator of the collective modes}
Eqs. \eqref{chi_general_norm}, \eqref{chi_general_anom}, and \eqref{fermion_propagator}
show that the calculation of the inverse propagator, $\Xi$, of the bosonic collective
excitations can be split into three independent steps. First, the trace over
Dirac indices is always of the form
$$
\tr_{\text D}(\Lambda_{\vec k}^e\Lambda_{\vec q}^f)=1+ef\frac{m^2+\vec k\cdot\vec
q}{\eps_{\vec k}\eps_{\vec q}}.
$$
Second step is the trace over the flavor indices, $\tr_{\text F}$. This will be discussed
shortly. Last, the summation over the fermionic Matsubara frequencies $\omega_n$ gives
rise to functions $I^{ef}_{rs}(\vec k,\vec q;i\Omega_N)$ and $J^{ef}_{rs}(\vec k,\vec
q;i\Omega_N)$ defined in Appendix \ref{App:Matsubara}.

Putting all the pieces together, the normal and anomalous correlation
functions, $\chi^{\Delta\Delta^*}$ and $\chi^{\Delta\Delta}$, are given by the
formulas
\begin{widetext}
\begin{align}
\label{chi_master_norm}
\chi_{ab}^{\Delta\Delta^*}(i\Omega_N,\vec
p)&=\frac1G\delta_{ab}+\frac12\sum_{r,s}\tr_{\text F}(Q_a\mathcal P_r\he Q_b\mathcal P_s^*)
\sum_{e,f}
\int\frac{d^3\vec k}{(2\pi)^3}\tr_{\text D}(\Lambda_{\vec k}^e\Lambda_{\vec k
-\vec p}^f)I^{ef}_{rs}(\vec k,\vec k-\vec p;i\Omega_N),\\
\label{chi_master_anom}
\chi_{ab}^{\Delta\Delta}(i\Omega_N,\vec
p)&=\frac12\sum_{r,s}\tr_{\text F}(Q_a\mathcal P_rMQ_b\mathcal P_sM)\sum_{e,f}
\int\frac{d^3\vec k}{(2\pi)^3}\tr_{\text D}(\Lambda_{\vec k}^e\Lambda_{\vec k
-\vec p}^f)J^{ef}_{rs}(\vec k,\vec k-\vec p;i\Omega_N).
\end{align}
In particular, in the normal phase all $\Delta$'s are equal to zero so that
$\chi^{\Delta\Delta}=0$. Moreover, we have $\sum_{r,s}\tr_{\text F}(Q_a\mathcal P_r\he
Q_b\mathcal P_s^*)=\tr_{\text F}(Q_a\he Q_b)=N\delta_{ab}$, which can always be enforced
by a suitable choice of basis of the flavor algebra. $N$ is a normalization
factor which, in principle, can differ for different pairing channels (or, different
irreducible representations of the flavor symmetry). In the following, we will
assume that the fermions pair in a single channel, but the generalization is
obvious and straightforward. After this remark, we can write down the
correlation function in the normal phase as a special case of Eq.
\eqref{chi_master_norm},
\begin{multline*}
\chi(i\Omega_N,\vec p)=\frac1G+\frac N2\int\frac{d^3\vec k}{(2\pi)^3}\left\{
\left[1+\frac{m^2+\vec k\cdot(\vec k-\vec p)}{\eps_{\vec k}\eps_{\vec k-\vec
p}}\right]\left[\frac{2f(\eps_{\vec k}+\mu)-1}{i\Omega_N+2\mu+\eps_{\vec k}+\eps_{\vec k-\vec
p}}+\frac{1-2f(\eps_{\vec k}-\mu)}{i\Omega_N+2\mu-\eps_{\vec k}-\eps_{\vec k-\vec
p}}\right]\right.\\
\left.+2\left[1-\frac{m^2+\vec k\cdot(\vec k-\vec p)}{\eps_{\vec k}\eps_{\vec k-\vec
p}}\right]\frac{f(\eps_{\vec k}+\mu)-f(\eps_{\vec k-\vec q}-\mu)}
{i\Omega_N+2\mu+\eps_{\vec k}-\eps_{\vec k-\vec p}}\right\},
\end{multline*}
\end{widetext}
which is equivalent to Eq. (15) in Ref. \cite{Abuki:2006dv}.

Full information about the spectrum of bosonic collective states may be
obtained by investigating the spectral density, $\rho(\omega,\vec p)$,
associated to the matrix correlation function $\Xi$, defined by
\begin{equation}
\bigl[\Xi^{-1}(i\Omega_N,\vec
p)\bigr]_{ab}=-\int_{-\infty}^{+\infty}\frac{d\omega}\pi\frac{\rho_{ab}(\omega,\vec
p)}{i\Omega_N-\omega}.
\label{spectral_Xi_decomposition}
\end{equation}
In general, we would like to use the relation
\begin{equation}
\rho_{ab}(\omega,\vec p)=\im\bigl[\Xi^{-1}(\omega+i\delta,\vec p)\bigr]_{ab},
\label{def_rho}
\end{equation}
in order to extract the spectral information from the imaginary part of the
pair correlation function, analytically continued to the real axis. However,
this should be done with some care for Eq. \eqref{def_rho} holds only when the
spectral density $\rho_{ab}(\omega,\vec p)$ \emph{defined} by Eq.
\eqref{spectral_Xi_decomposition} is real.

\subsection{Mean-field analysis}
Assuming that in thermodynamic equilibrium the system relaxes to a homogeneous state, the
mean-field thermodynamic potential $\Omega_{\text{MF}}$ is given by Eq.
\eqref{Action_bosonized},
\begin{multline}
\frac{\Omega_{\text{MF}}}{V}=\frac{|\Delta_a|^2}G\\
-\sum_r\sum_{e=\pm}\int\frac{d^3\vec
k}{(2\pi)^3}\bigl[(E_{\vec kr}^e-\eps_{\vec k}+2T\log(1+e^{-\beta E_{\vec
kr}^e})\bigr],
\label{MF_TDpot}
\end{multline}
where we subtracted the energy of the vacuum in order that $\Omega_{\text{MF}}$ expresses
the sheer effect of finite temperature and density, and the Cooper pairing.

It should be noted that Eq. \eqref{MF_TDpot} embodies two different sets of gap
parameters. First, the $\Delta_a$'s, which carry the label of the Cooper pair
representation of the symmetry, and represent the order parameters for symmetry breaking. 
Second, the $\Delta_r$'s, which determine the gaps in the fermion excitation spectrum. 
While the $\Delta_a$'s transform as an antisymmetric tensor under the symmetry
transformations acting on $\psi$ and may in general be complex, the $\Delta_r$'s are by
construction invariants of the symmetry and are always real. 

These two sets of parameters may be related with the
help of the expression $\Delta_a\Delta^*_b\he Q_aQ_b=M\he M=\sum_r\Delta_r^2\mathcal P_r$.
By taking the trace we find
$$
N\sum_a|\Delta_a|^2=\sum_r\Delta_r^2.
$$
This clarifies the physical content of the normalization factor $N$: It counts the number
of fermion degrees of freedom participating in the formation of a particular Cooper pair.
Plugging this result back into Eq. \eqref{MF_TDpot}, we immediately arrive at the
standard mean-field gap equation and the equation for the particle number density $n$,
\begin{align}
\label{MF_gapeq}
\Delta_r&=\Delta_rNG\sum_{e=\pm}\int\frac{d^3\vec k}{(2\pi)^3}\frac{1-2f(E_{\vec
kr}^e)}{2E_{\vec kr}^e},\\
\label{MF_numeq}
n&=\sum_r\sum_{e=\pm}\int\frac{d^3\vec k}{(2\pi)^3}\frac{e\xi_{\vec k}^e}{E_{\vec
kr}^e}\left[1-2f(E_{\vec kr}^e)\right].
\end{align}
The total number density $n$ is fixed by the value of the Fermi momentum $k_{\text F}$,
as $n=k_{\text F}^3/(3\pi^2)$ times the number of fermion flavors.

\subsection{Renormalization}
\label{Sec:renormalization}
The momentum integral in Eq. \eqref{MF_gapeq} (as well as in other equations depending
explicitly on the coupling) is badly divergent. This can be taken care of by trading the
bare coupling $G$ for the physical $s$-wave scattering length at zero temperature and
density, or the renormalized coupling $G_R$, defined by \cite{Abuki:2006dv}
$$
\frac1G=-\frac1{G_R}+\frac N2\int\frac{d^3\vec k}{(2\pi)^3}\left(\frac1{\eps_{\vec
k}+m}+\frac1{\eps_{\vec k}-m}\right).
$$
As opposed to the case of nonrelativistic fermion matter, the remaining
momentum integration is still divergent, but only mildly, logarithmically. In
numerical computations, it is regulated with a sharp three-momentum cutoff
$\Lambda$. All equations of the theory may be rewritten in such a way that the
cancelation of the leading divergences is manifest. The resulting formulas are
useful for the numerical implementation, but otherwise are rather cumbersome.
Just for illustration, we show here the gap equation and the equation for number density,
\begin{multline*}
\Delta_r=-\frac12\Delta_rNG_R\sum_{e=\pm}\int\frac{d^3\vec
k}{(2\pi)^3}\left[-\frac{m^2}{\vec k^2\eps_{\vec k}}-\frac{2f(E^e_{\vec kr})}
{E^e_{\vec kr}}\right.\\
\left.+\frac{2\mu^2}{E^e_{\vec kr}E^{-e}_{\vec kr}(E^e_{\vec kr}+E^{-e}_{\vec kr})}
-\frac{\Delta_r^2}{\eps_{\vec k}}\frac1{E^e_{\vec kr}(E^e_{\vec kr}+\xi_{\vec k}^e)}
\right],
\end{multline*}
\begin{multline*}
n=\sum_r\sum_{e=\pm}\int\frac{d^3\vec k}{(2\pi)^3}\left[-2e\frac{\xi_{\vec k}^e}{E^e_{\vec
kr}}f(E^e_{\vec kr})\right.\\
\left.+\frac{\mu\Delta_r^2}{E^e_{\vec kr}E^{-e}_{\vec kr}(E^e_{\vec kr}+E^{-e}_{\vec kr})}
\left(1+\frac{E^e_{\vec kr}+\xi_{\vec k}^e}{E^{-e}_{\vec kr}+\xi_{\vec
k}^{-e}}\right)\right].
\end{multline*}

The procedure outlined above is sufficient to renormalize the mean-field
thermodynamic potential and its derivates, the gap and number equations. On
the other hand, one must be more careful when calculating the correlator
$\Xi_{ab}$ \eqref{Xi_matrix}, or in general any fermion loop with nonzero
external momentum. The point is that the momentum assignment for the internal
fermion lines is not completely fixed by momentum conservation in the
interaction vertices; it is unique up to a shift of the integration variable.
Such a shift, however, is not permitted in an integral with stronger than
logarithmic divergence \footnote{Suppose that we integrate in $D$ dimensions a function
$f(\vec k)$ with asymptotics $f(\vec k)\sim1/|\vec k|^n$ as $|\vec k|\to\infty$, with
integer $n$. A sharp cutoff restricts the integration domain to a sphere of radius
$\Lambda$. Shifting the integration variable by a small vector $\vec a$ is equivalent to
changing the integration domain by a thin shell (of size $\sim|\vec a|$). The
corresponding change of the integral is estimated by the value of the function on the
shell times its volume, i.e., $(1/\Lambda^n)\Lambda^{D-1}|\vec a|=|\vec
a|\Lambda^{D-n-1}$. This will be negligible (and the shift thus permitted) provided $n\geq
D$.}. Therefore, we are led to consider the momentum assignment to the fermion propagators
in Eqs. \eqref{chi_general_norm} and \eqref{chi_general_anom} as a part of the
\emph{definition} of the model. As already explained above, the integrals are then
evaluated with a sharp cutoff on the three-momentum $\vec k$.

Let us finally remark that one might at first sight think it would be natural
to impose the cutoff directly on the arguments of the fermion propagators in
the loop, i.e., on both $\vec k$ and $\vec k-\vec p$ in Eqs.
\eqref{chi_general_norm} and \eqref{chi_general_anom}; this would roughly
correspond to discretizing the coordinate space. Nevertheless, this would
destroy the expected low-energy dynamics of our model, based on the
Nambu--Goldstone (NG) modes of the spontaneously broken symmetry --- the
inverse propagator \eqref{chi_general_norm} would develop a linearly divergent
piece, proportional to $|\vec p|$.

\section{Two-flavor color superconductor}
Now we apply the general formulas derived in the previous section to the
particular model studied by Abuki \cite{Abuki:2006dv}. It describes color
superconductivity in quark matter consisting of two quark flavors, which have
for simplicity equal masses and chemical potentials. The field $\psi$ now
carries the index of the fundamental representation of the symmetry group
$\mathrm{SU(3)\times SU(2)}$. Since the weak-coupling studies of quark matter
indicate attraction between two quarks in the color-antitriplet channel, we
assume that the matrices $Q_a$ have the structure
$(Q_a)^{ij}_{bc}=\eps^{ij}\eps_{abc}$, where $i,j$ from now on denote the
flavor $\mathrm{SU(2)}$ indices and $a,b,c$ the color $\mathrm{SU(3)}$ ones.
The order parameter $\Delta_a$ thus transforms as an $\mathrm{SU(2)}$-singlet
and an $\mathrm{SU(3)}$-antitriplet. As usual it is chosen to be real and to
point in the third (anti-blue) direction in the color space. In the following,
we write simply $\Delta$ instead of $\Delta_3$; this will help us distinguish
the constant order parameter from the fluctuation fields without having to
introduce further notation.

For this particular symmetry structure and the choice of vacuum, we find $N=4$
and $M\he M=\Delta^2\mathrm{diag}(1,1,0)$ in the color space (in the flavor
space, it is simply the unit matrix). So in this case, the projectors $\mathcal P_r$
simply project on the individual colors. The red and green quarks,
participating in the pairing, are gapped, while the blue quarks remain gapless.
As is well known, the global $\mathrm{SU(3)\times U(1)}$ symmetry of the
bosonized action \eqref{Action_bosonized} is broken by the Cooper pair
condensate $\Delta$ down to $\mathrm{SU(2)\times U(1)_Q}$, operating
exclusively on $\Delta_1$ and $\Delta_2$ \footnote{The scalar field $\Delta_a$
is a singlet of the original flavor $\mathrm{SU(2)}$ symmetry so that this is
no more a symmetry of the bosonized action \eqref{Action_bosonized}. The flavor
quantum number has been ``integrated out'' together with the quarks.}. The
unbroken symmetry implies degeneracy between $\Delta_1$ and $\Delta_2$. The
pair correlation matrix $\Xi_{ab}$ is then diagonal and $\Xi_{11}=\Xi_{22}$.
Moreover, $\Xi_{11}$ has only the normal part, thanks to the existence of the
unbroken $\mathrm{U(1)_Q}$ charge. Evaluating explicitly the color--flavor
traces in Eqs. \eqref{chi_master_norm} and \eqref{chi_master_anom}, we obtain
the expressions
\begin{widetext}
\begin{align*}
\chi_{11}^{\Delta\Delta^*}(i\Omega_N,\vec
p)&=\frac1G+\sum_{e,f}\int\frac{d^3\vec k}{(2\pi)^3}\left[1+ef\frac{m^2+\vec
k\cdot(\vec k-\vec p)}{\eps_{\vec k}\eps_{\vec k-\vec
p}}\right][I^{ef}_{23}(\vec k,\vec k-\vec
p;i\Omega_N)+I^{ef}_{32}(\vec k,\vec k-\vec p;i\Omega_N)],\\
\chi_{33}^{\Delta\Delta^*}(i\Omega_N,\vec
p)&=\frac1G+2\sum_{e,f}\int\frac{d^3\vec k}{(2\pi)^3}\left[1+ef\frac{m^2+\vec
k\cdot(\vec k-\vec p)}{\eps_{\vec k}\eps_{\vec k-\vec
p}}\right]I^{ef}_{12}(\vec k,\vec k-\vec
p;i\Omega_N),\\
\chi_{33}^{\Delta\Delta}(i\Omega_N,\vec
p)&=-2\Delta^2\sum_{e,f} \int\frac{d^3\vec k}{(2\pi)^3}\left[1+ef\frac{m^2+\vec
k\cdot(\vec k-\vec p)}{\eps_{\vec k}\eps_{\vec k-\vec
p}}\right]J^{ef}_{12}(\vec k,\vec k-\vec p;i\Omega_N).
\end{align*}
\end{widetext}
Note that the quark loop contributing to $\Xi_{33}$ is symmetric: Both
propagators correspond to gapped quarks. On the other hand, the loop in
$\Xi_{11}$ consists of one gapped and one ungapped propagator and the general
formula \eqref{chi_master_norm} requires it to be symmetrized.

\subsection{Nambu--Goldstone bosons}
Spontaneous symmetry breaking leads to the existence of NG bosons, soft
fluctuations of the order parameter. Vacuum expectation value of a color
antitriplet breaks five generators of the color $\mathrm{SU(3)}$ so that we
should expect five NG bosons. However, in our model where the color symmetry is
just global unlike in quantum chromodynamics, the Noether charge corresponding
to the Gell-Mann matrix $\lambda_8$ acquires nonzero density. This implies that
there are only three NG bosons, two of which have quadratic dispersion law at
low momentum, being the so-called type-II NG bosons
\cite{Nielsen:1976hm,Blaschke:2004cs,Brauner:2007uw}. These couple to the two pairs of
generators, $(\lambda_4,\lambda_5)$ and $(\lambda_6,\lambda_7)$, whose commutator
contains a $\lambda_8$ piece and hence has nonzero density. Therefore, they will be found
in the spectrum of excitations of $\Delta_1$ and $\Delta_2$. The remaining, type-I NG
boson couples to the spontaneously broken linear combination of $\lambda_3$ and
$\lambda_8$ and will be found as an excitation of $\Delta_3$.

At zero temperature, the presence of NG modes in our model may easily be demonstrated
analytically. We set $\vec p=\vec0$, use the gap equation \eqref{MF_gapeq}, and
analytically continue the pair correlation functions to real frequency (or, more
precisely, to the vicinity of the real axis),
\begin{multline*}
\chi_{11}^{\Delta\Delta^*}(\omega,\vec0)=2\omega\int\frac{d^3\vec k}{(2\pi)^3}\left\{
\frac1{E_{\vec k}^+}\frac1{\omega+E_{\vec k}^++\xi_{\vec k}^+}\right.\\
\left.+\frac1{E_{\vec k}^-}\left[\frac{\theta(\xi_{\vec k}^-)}{\omega-E_{\vec
k}^--|\xi_{\vec k}^-|}+\frac{\theta(-\xi_{\vec k}^-)}{\omega+E_{\vec k}^-+|\xi_{\vec
k}^-|}\right]\right\},
\end{multline*}
\begin{align*}
\chi_{33}^{\Delta\Delta^*}(\omega,\vec0)&=2\sum_{e=\pm}\int\frac{d^3\vec
k}{(2\pi)^3}\frac1{E_{\vec k}^e}\frac{\omega^2-2\Delta^2-2e\omega
\xi_{\vec k}^e}{\omega^2-(2E_{\vec k}^e)^2},\\
\chi_{33}^{\Delta\Delta}(\omega,\vec0)&=4\Delta^2\sum_{e=\pm}\int\frac{d^3\vec
k}{(2\pi)^3}\frac1{E_{\vec k}^e}\frac{1}{\omega^2-(2E_{\vec k}^e)^2}.
\end{align*}
[We have dropped the index $r$ so that $E^e_{\vec k}$ is now defined by Eq.
\eqref{E_xi_definition} with the gap $\Delta$ inserted.] We can immediately see
that $\chi_{11}^{\Delta\Delta^*}(\omega,\vec0)\to0$ as $\omega\to0$, which
proves the existence of NG excitations coupled to $\Delta_1$ and $\Delta_2$.
These are therefore interpreted as massless bound states of one gapped and one
gapless quark (green and blue, and red and blue, respectively).

The correlation functions of $\Delta_3$ do not go to zero at vanishing
frequency. However, here the propagator is given by inverting the matrix $\Xi$
\eqref{Xi_matrix}. Since in the limit $\omega\to0$ we have
$$
\chi_{33}^{\Delta\Delta^*}(0,\vec0)=-\chi_{33}^{\Delta\Delta}(0,\vec0),
$$
the determinant of $\Xi$ will be zero and the existence of a NG mode is thus proved.

At zero temperature, we can of course go beyond the mere demonstration of the
existence of NG bosons, and determine their low-momentum dispersion relation.
This may be done in the standard manner, by expanding the correlation functions
in powers of momentum. Let us just remark that at nonzero temperature but still
in the superfluid phase, there are kinematical regions where the sharp NG peak
in the pair correlation spectrum is obscured by Landau damping --- the emission
or absorption of the NG boson by a quark. In such regions the momentum
expansion of the correlation functions is not possible \cite{Abrahams:1966ab}.

Here we concentrate on the doublet of type-II NG bosons coupled to
$\Delta_{1,2}$, aiming to investigate the effect of the expected quadratic
onset of their dispersion relation. At low frequency and momentum, the inverse
propagator $\chi^{\Delta\Delta^*}_{11}$ is expanded as
\begin{equation}
\chi_{11}^{\Delta\Delta^*}(\omega,\vec p)=-a\omega+b\vec p^2.
\label{ab_coefficients}
\end{equation}
The coefficients $a,b$ determine the low-momentum dispersion relation of the NG
boson by $\omega(\vec p)=\vec p^2b/a$. They are given by the explicit
expressions
$$
a=\frac2{\Delta^2}\sum_{e=\pm}\int\frac{d^3\vec
k}{(2\pi)^3}\left[\frac{e\xi^e_{\vec k}}{E^e_{\vec k}}-\sgn(e\xi^e_{\vec
k})\right],
$$
\begin{multline*}
b=\theta(\mu-m)\frac{(\mu^2-m^2)^{3/2}}{3\pi^2\Delta^2\mu}
+\sum_{e=\pm}\int\frac{d^3\vec k}{(2\pi)^3}\\
\times\left[\left(1-\frac{\vec k^2}{3\eps_{\vec k}^2}\right)\frac1{\eps_{\vec
k}E^e_{\vec k}}\left(\frac{\sgn\xi^{-e}_{\vec k}}{E^e_{\vec k}+|\xi^{-e}_{\vec
k}|}+\frac{\xi^e_{\vec k}}{2(E^e_{\vec k})^2}\right)\right.\\
-\left.\frac{2\vec k^2}{3\eps_{\vec k}^2}\frac1{E^e_{\vec k}}\frac1{(E^e_{\vec
k}+|\xi^e_{\vec k}|)^2}+\frac{\vec k^2}{2\eps_{\vec k}^2}\frac{\Delta^2}{(E^e_{\vec
k})^5}\right]
\end{multline*}
Note that $a$ is actually equal to $1/\Delta^2$ times the difference of the
density of a gapped and a gapless  color [cf. Eq. \eqref{MF_numeq}, the factor $2$ arises
from the existence of two quark flavors for each color]. This in turn is
proportional to the density of $\lambda_8$ as it should, for it is this
non-Abelian charge which, according to general theorems
\cite{Schaefer:2001bq,Brauner:2007uw} gives rise to type-II NG bosons.
\begin{figure}[t]
\includegraphics[scale=0.45]{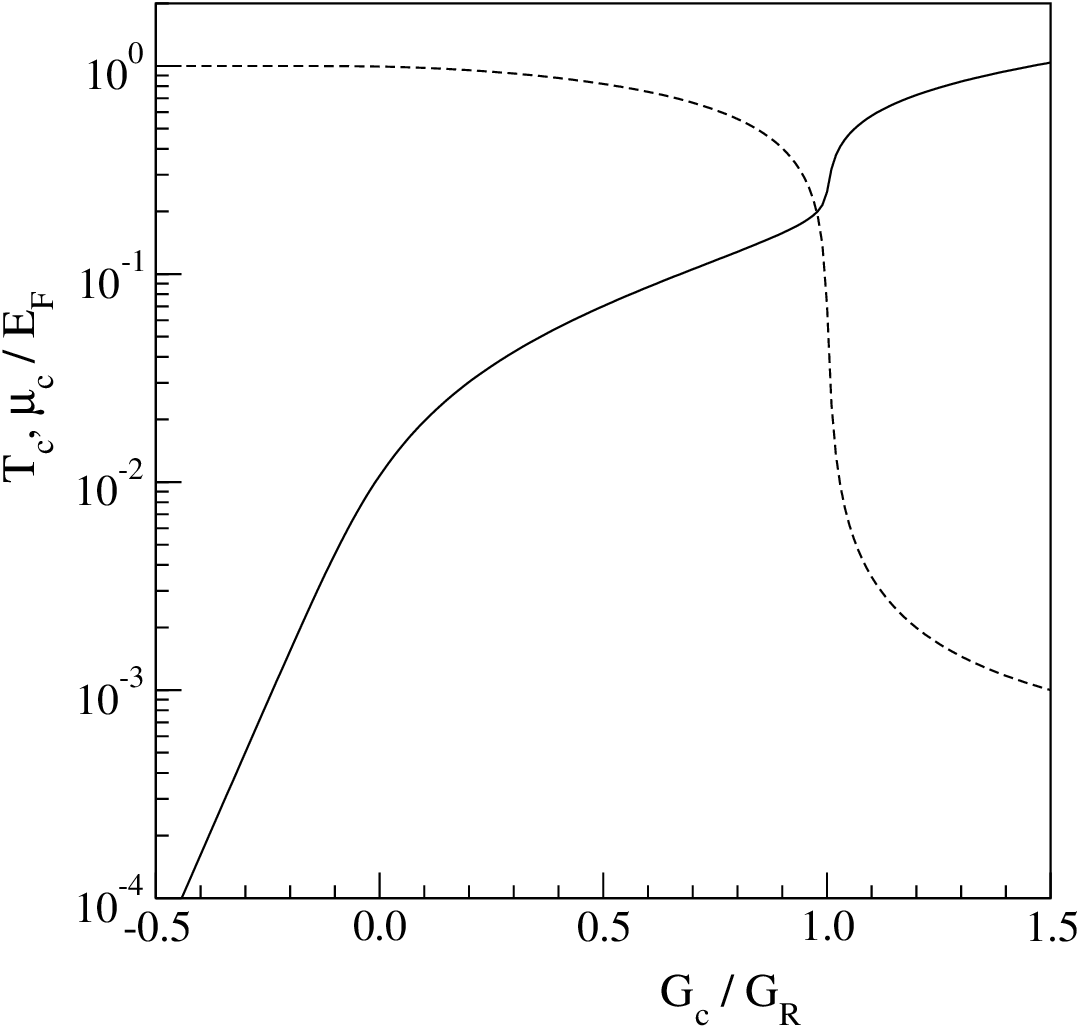}
\caption{The critical temperature (solid) and chemical potential (dashed) in
units of the Fermi energy, as a function of the inverse coupling. Three different regimes
are distinguished: BCS ($G_c/G_R\lesssim0$), BEC ($0\lesssim G_c/G_R\lesssim1$), and RBEC
($G_c/G_R\gtrsim1$).}
\label{Fig:phase_diagram}
\end{figure}

\subsection{Numerical results}
In this section, we present numerical results that illustrate the general
conclusions made above. In order to achieve concrete numbers, we use the same
set of parameters as Abuki \cite{Abuki:2006dv} so that our results in the
superfluid phase can be matched to his for the normal phase. In particular, we
choose the Fermi momentum $k_{\text F}=0.2m$ and a cutoff to regulate the
logarithmic-divergent integrals as $\Lambda=5m$. The renormalized coupling
$G_R$ is represented in units of the critical coupling $G_c$, at which the mass
of the diquark molecule in vacuum goes to zero. This signifies the instability
of the vacuum itself with respect to pair formation, and the onset of
relativistic Bose--Einstein condensation (RBEC).

\subsubsection{Phase diagram}
The dependence of the critical temperature on the renormalized coupling is
displayed in Fig. \ref{Fig:phase_diagram}. It is instructive to compare this
mean-field plot to a similar one achieved by Abuki and Nishida
\cite{Nishida:2005ds,Abuki:2006dv}, who included the effects of Gaussian
fluctuations. (To be precise, they combine the mean-field Thouless criterion
\cite{Thouless:1960th} with the equation for particle number that includes the
contribution of bosonic quasiparticles.) It is clearly seen that as expected,
the mean-field approximation is reliable in the BCS regime (negative coupling),
while it fails by about an order of magnitude on the BEC side. The reason is,
of course, that the mean-field thermodynamic potential includes only the
contribution of the fermions, whereas in the BEC regime the total number
density is dominated by the bosons. In the RBEC limit (that is for
$G_c/G_R\gtrsim1$), both bosons and fermions are excited significantly, and the
reliability of the mean-field approximation again improves.

\begin{figure}[t]
\includegraphics[scale=0.45]{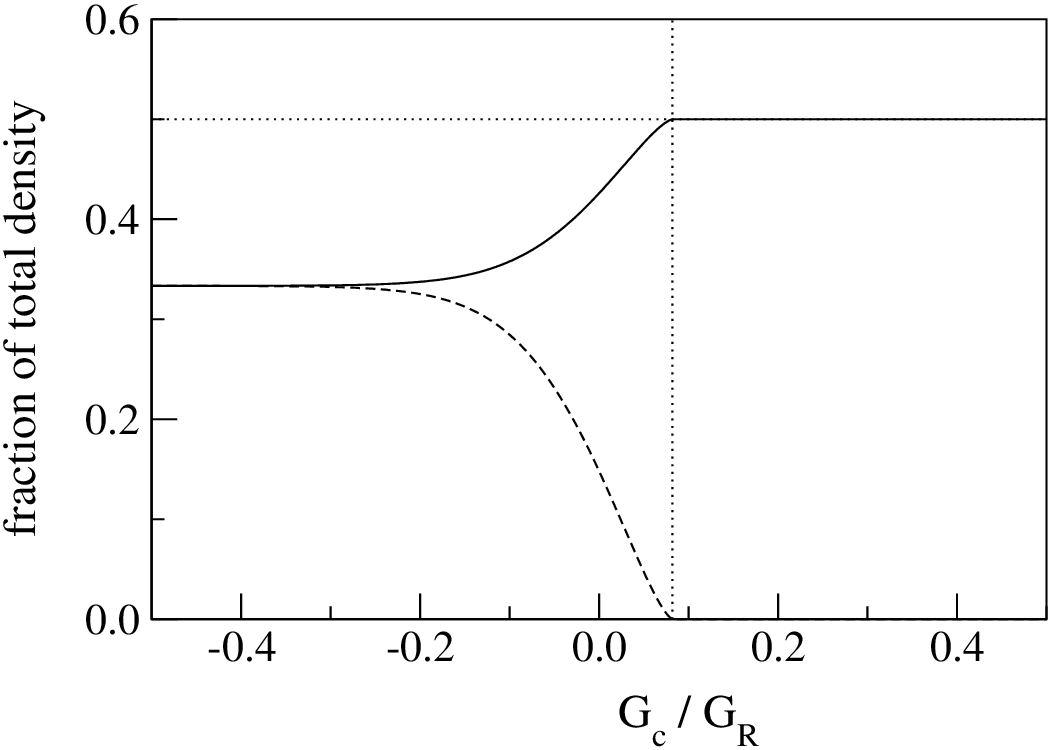}
\caption{Zero-temperature density of quarks of a single paired (solid) and
unpaired (dashed) color at zero temperature, expressed as a fraction of the
fixed total density. Recall that there are two paired colors (red and green)
and one unpaired one (blue).} \label{Fig:density}
\end{figure}

In Fig. \ref{Fig:density} we show the contributions of the paired and unpaired
quarks to the total density at zero temperature. In the BCS limit the unpaired
blue quarks form a sharp Fermi sea, while the paired red and green quarks
occupy a Fermi sea, smeared about the Fermi surface at the scale of the gap.
However, at $G_c/G_R\approx0.08$ the chemical potential decreases below the
fermion mass and the Fermi sea can no longer exist. This marks the onset of the
BEC regime. From this point on, the whole density at zero temperature is
provided by the condensate of the bound diquark molecules, made of red and
green quarks. The blue quarks can only be excited thermally.

\subsubsection{Bosonic spectrum at zero temperature and momentum}
As already remarked before, the correlation function $\Xi_{ab}$ is diagonal in
the color--flavor space, owing to the choice of the order parameter $\Delta$ to
point in the third direction in the anticolor space. Moreover, since we choose
$\Delta$ real, the spectral density $\rho_{ab}$ defined by Eq.
\eqref{spectral_Xi_decomposition} is real and we can use Eq. \eqref{def_rho} to
calculate it.

\begin{figure}[t]
\includegraphics[scale=0.45]{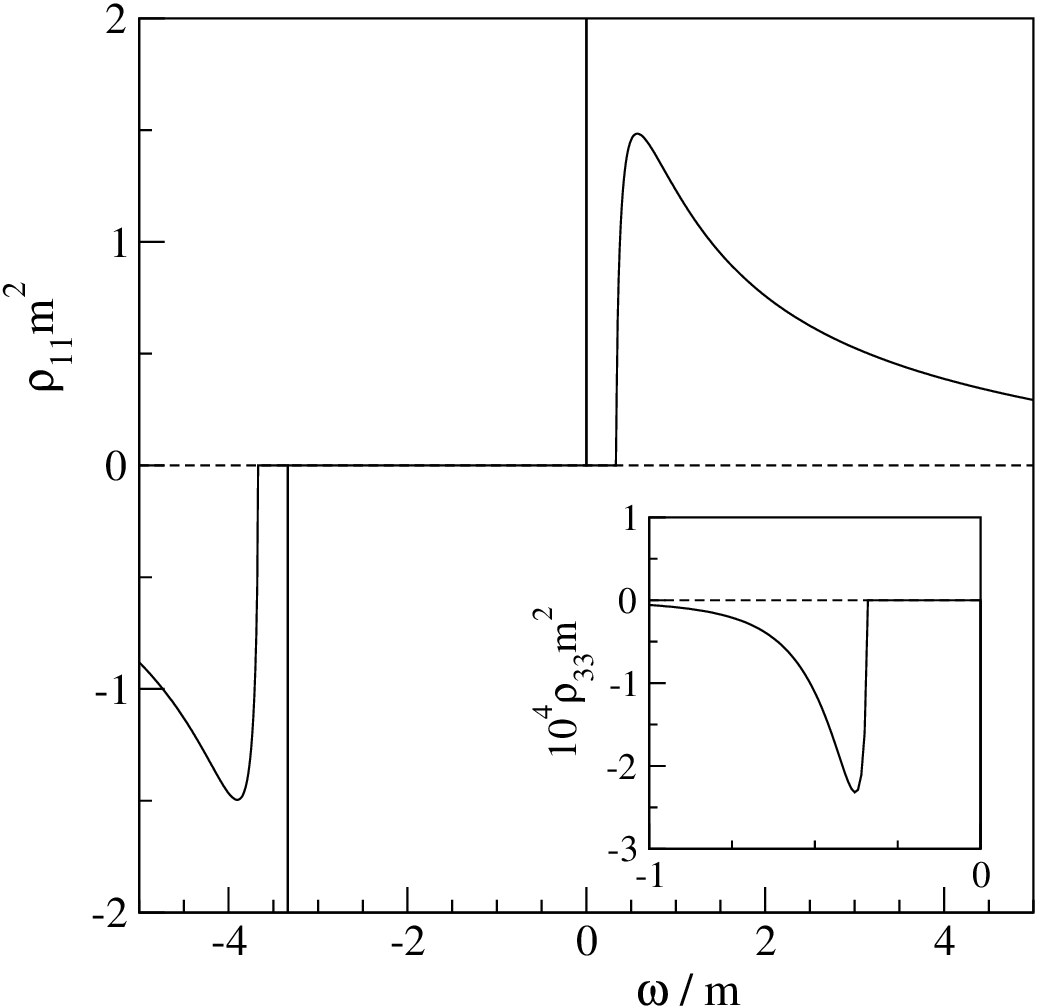}
\caption{Normal component of the spectral density matrix at zero temperature
and momentum for $G_c/G_R=0.5$. The large-scale plot applies equally well to
$\rho_{11}$ and $\rho_{33}$. In the inset we show the tiny contribution of the
particle--particle continuum at negative frequencies to $\rho_{33}$, which is
absent in $\rho_{11}$.} \label{Fig:rho_at_T0}
\end{figure}

In the normal phase, the matrix $\Xi$ is simply proportional to the unit matrix
and there is a single spectral density, which has been calculated at the
critical temperature by Abuki \cite{Abuki:2006dv}. In the superfluid phase, we
have one spectral density for $a=b=1,2$ and a symmetric $2\times2$ matrix for $a=b=3$. In
Fig. \ref{Fig:rho_at_T0} we give a sample numerical calculation of the normal
parts of these spectral densities for $G_c/G_R=0.5$.

Several comments are in order here. First, in the normal phase an isolated
bound state (corresponding to a diquark or antidiquark molecule) exists only on
the BEC side of the crossover. In the superfluid phase, the NG bound state is
present for all values of the coupling, as guaranteed by the Goldstone theorem.
On the other hand, the antiboson pole appears only on the BEC side of the
crossover, just as in the normal phase. It should be stressed that the
positions of the antiboson poles in $\rho_{11}$ and $\rho_{33}$ differ, simply
because one of the bosons is composed of a gapped and an ungapped (anti)quark, while
the other one of two gapped (anti)quarks. Indeed, the pole in $\rho_{33}$ occurs at a
smaller (negative) $\omega$; the corresponding boson is heavier. Nevertheless,
for $G_c/G_R=0.5$ the difference is so small that it cannot be seen in Fig.
\ref{Fig:rho_at_T0}.

Apart from the presence of the isolated antiboson pole on the BEC side of the
crossover (and its absence on the BCS side), the qualitative appearance of the
spectrum does not change as the coupling is varied. As noted above, the
antiboson pole has different positions in $\rho_{11}$ and $\rho_{33}$. The
same, of course, holds for the onset of the continuum in these spectra. As we
proceed from the BCS to the BEC to the RBEC regime, the gap in the spectrum of
red and green quarks increases, and the difference between $\rho_{11}$ and
$\rho_{33}$ becomes more pronounced.

\begin{figure}[t]
\includegraphics[scale=0.45]{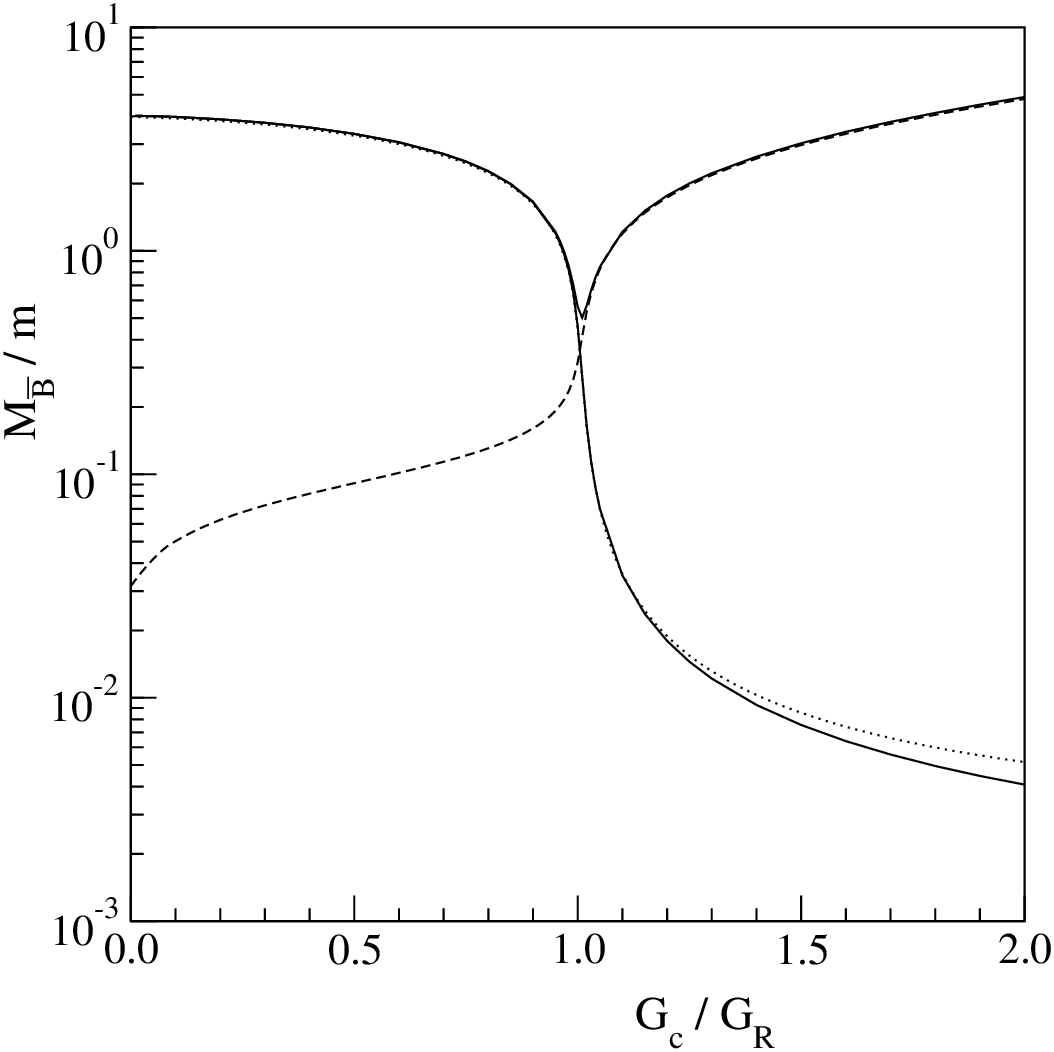}
\caption{Absolute value of the position of the antiboson pole in $\rho_{11}$
(lower solid line) and $\rho_{33}$ (upper solid line) at zero temperature and
momentum. For comparison, we also show the values of $2\Delta$ (dashed) and
$4\mu$ (dotted).} \label{Fig:antiboson_mass}
\end{figure}

In Fig. \ref{Fig:antiboson_mass} we show the coupling-dependence of the two
antiboson masses. In the BEC regime, they are almost equal and essentially
determined by the chemical potential. However, this changes dramatically in the
RBEC regime. While the pole in $\rho_{33}$ appears near the two-antiparticle
continuum, now dominated by the large value of the gap, the antiboson pole in
$\rho_{11}$ is still driven by the chemical potential. We checked numerically that as the
temperature is increased, the mass of the antiboson in $\rho_{33}$
follows (twice) the value of the gap until very close to the phase transition, where it
merges with the mass of the antiboson in $\rho_{11}$. This
is in agreement with Ref. \cite{Abuki:2006dv} where it was shown that in the
normal phase, both masses (being equal) are determined by the chemical potential.

Another fact to note is the order-of-magnitude suppression of the anomalous
spectral density $\rho_{33}^{\Delta\Delta}$. This is due to the fact that, with
the exception of the RBEC limit, the gap is much smaller than other scales in
the theory such as the Fermi energy or chemical potential. This tiny
off-diagonal (in the Nambu space) element of the correlation function matrix
$\Xi$ then induces mixing between the bound boson and antiboson states, or in
other words, breaks the conservation of the baryon number carried by the
quarks. Consequently, the spectral density $\rho_{33}^{\Delta\Delta^*}$
displays, besides the expected two-particle continuum at positive frequency,
also an analogous continuum at negative frequency, yet with much smaller
spectral weight. In $\rho_{11}^{\Delta\Delta^*}$, this continuum is absent due
to the existence of the conserved charge of the unbroken $\mathrm{U(1)_Q}$.

\subsubsection{Bosonic spectrum at nonzero temperature and momentum}
\begin{figure*}[t]
\includegraphics[scale=0.7]{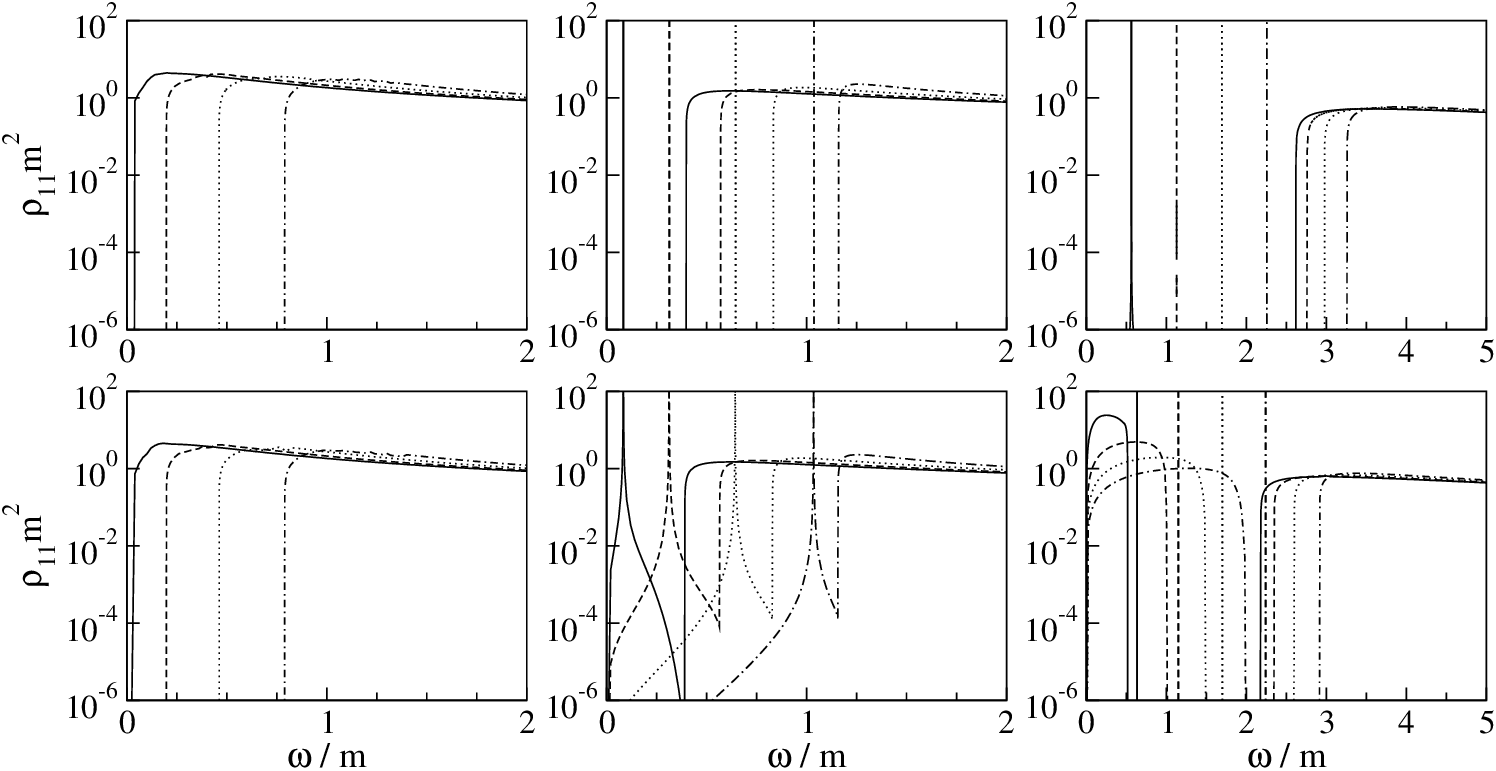}
\caption{Positive-frequency part of the $\rho_{11}$ spectral density at nonzero
temperature and momentum. First column: BCS regime, $G_c/G_R=-0.35$; second
column: BEC regime, $G_c/G_R=0.5$; third column: RBEC regime, $G_c/G_R=1.35$.
The upper line shows spectra at low temperature, $T=0.05T_c$, the lower line
near the phase transition, $T=0.95T_c$. The four curves in each spectrum
correspond respectively to $|\vec p|=0.5m$ (solid), $|\vec p|=1.0m$ (dashed),
$|\vec p|=1.5m$ (dotted), and $|\vec p|=2.0m$ (dash-dotted).}
\label{Fig:GB_spectra_nonzero_T}
\end{figure*}

At nonzero temperature and momentum the spectrum becomes more rich. The
bound-state peaks acquire nonzero (even though tiny) width and the Landau
damping appears. Having in mind that the low-energy dynamics of the system in
the superfluid phase is dominated by the NG bosons, we display in Fig.
\ref{Fig:GB_spectra_nonzero_T} the positive-frequency parts of the spectral
density $\rho_{11}$; each panel collects the spectra for several values of the
momentum. The NG excitations are manifested by isolated peaks below the
two-particle continuum.
Comparison of spectra for temperatures close to zero and to the phase
transition reveals the temperature effects: The overall increase of the
background due to thermal excitations, and the Landau damping, especially
strong in the RBEC regime. We can also see how with increasing momentum the NG
boson peaks get closer to the continuum; at a certain point they will reach the
continuum and disappear from the spectrum as isolated poles.

\subsubsection{NG boson dispersion relations}
\begin{figure*}[t]
\begin{center}
\vskip2ex
\includegraphics[scale=0.65]{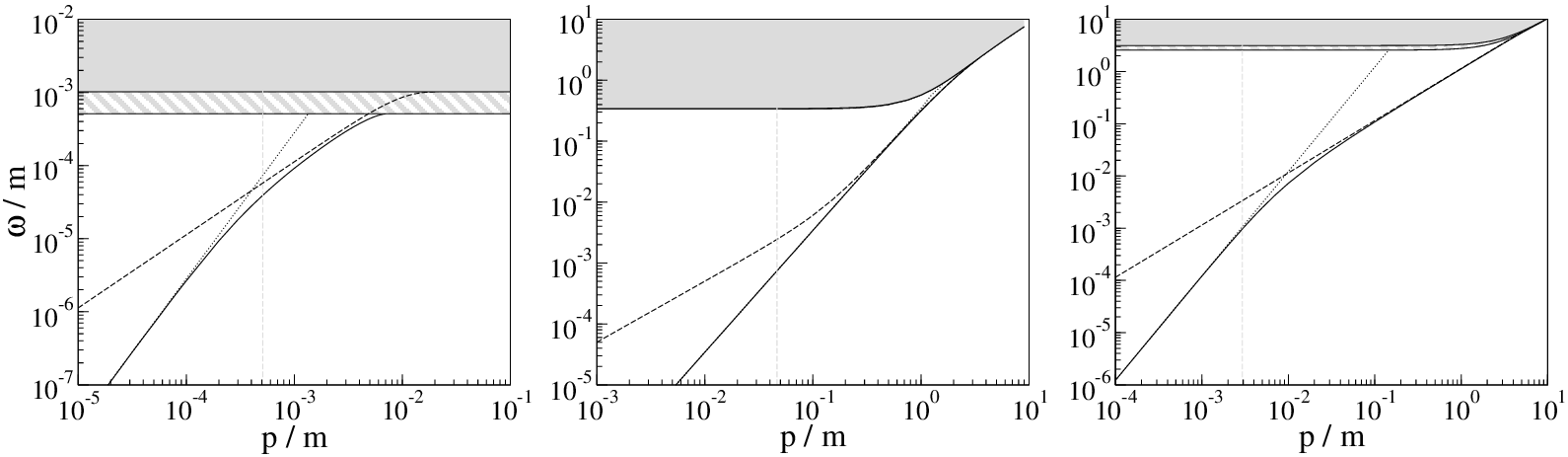}
\end{center}
\caption{Dispersion relations of the NG bosons. First plot: BCS regime,
$G_c/G_R=-0.35$; second plot: BEC regime, $G_c/G_R=0.5$; third plot: RBEC
regime, $G_c/G_R=1.35$. In all graphs, the dispersion relation of the type-II
NG boson in $\rho_{11}$ (solid), its low-momentum prediction (dotted), as well
as the dispersion relation of the type-I NG boson in $\rho_{33}$ (dashed) are
shown. The hatched-gray and solid-gray regions denote the continuum in $\rho_{11}$ and
$\rho_{33}$, respectively. The dashed gray line indicates the value of the gap (BCS,
BEC), or the chemical potential (RBEC).}
\label{Fig:GB_dispersions}
\end{figure*}
In addition to the full spectra, we also studied the dispersion relations of
the NG bosons at zero temperature. These can be extracted from the spectral
densities and, in case of the type-II NG bosons, compared to the analytical
low-momentum prediction from Eq. \eqref{ab_coefficients}. The result is shown
in Fig. \ref{Fig:GB_dispersions}.

General consequences of spontaneous symmetry breaking in presence of nonzero
charge density \cite{Brauner:2007uw} require that the dispersion relation of
the NG boson in $\rho_{11}$ and $\rho_{33}$ is, at low momentum, quadratic and
linear, respectively. The scale that defines this low-momentum region is
typically set by the symmetry breaking. Above this scale, the NG boson
dispersion relation is no longer fixed by the general properties of symmetry
breaking, but rather is determined by the detailed dynamics of the system under
consideration. In some cases, the NG boson may even cease to exist at high
enough momentum; in our model, this is reflected by its disappearance in the
continuum.

We display the dispersion relations in a logarithmic scale in order to exhibit
their power-law behavior. Apparently, the low-momentum region governed by the
broken symmetry is indeed defined by the symmetry-breaking scale, i.e., the gap.
The only exception is the RBEC limit where the predicted quadratic dispersion
of the type-II NG boson is observed only below the scale of the chemical
potential, which is much smaller than the gap. This is because above the
chemical potential the particles and antiparticles become nearly degenerate and
all effects of finite density are strongly suppressed.

In the BEC regime, we can even distinguish three regions of momentum with
physically distinct behavior. First, at $|\vec p|\lesssim\Delta$, the
dispersions of the NG bosons are governed by the broken symmetry and match the
classification into type I and type II. Second, for momenta larger than a few times the
gap but smaller than the fermion mass $m$, the dispersions of all NG bosons
become quadratic. This is in accord with the Bogolyubov theory of nonrelativistic BEC
\cite{Andersen:2003qj}. Finally, for $|\vec p|\gtrsim m$ we enter the relativistic regime
and the dispersions become linear again. (Although it cannot be seen in Fig.
\ref{Fig:GB_dispersions}, the NG bosons disappear in the continuum at $|\vec
p|\approx3.6m$. So in a linear scale, this relativistic domain actually covers a large
part of the dispersion relations.)

Let us also remark that in the BCS regime, the frequency marking the onset of the
two-particle continuum remains strictly constant for all values of momentum displayed in
Fig. \ref{Fig:GB_dispersions}. The reason is that only in the BCS regime has the
dispersion of the fermionic quasiparticles a nontrivial minimum. Consequently, for all
momenta $|\vec p|$ up to the Fermi momentum, $k_F$, the pair mode may decay into two
fermions both lying on their Fermi surfaces. So the continuum sets at $\omega=\Delta$ in
$\rho_{11}$ and at $\omega=2\Delta$ in $\rho_{33}$.

\begin{figure}[t]
\includegraphics[scale=0.45]{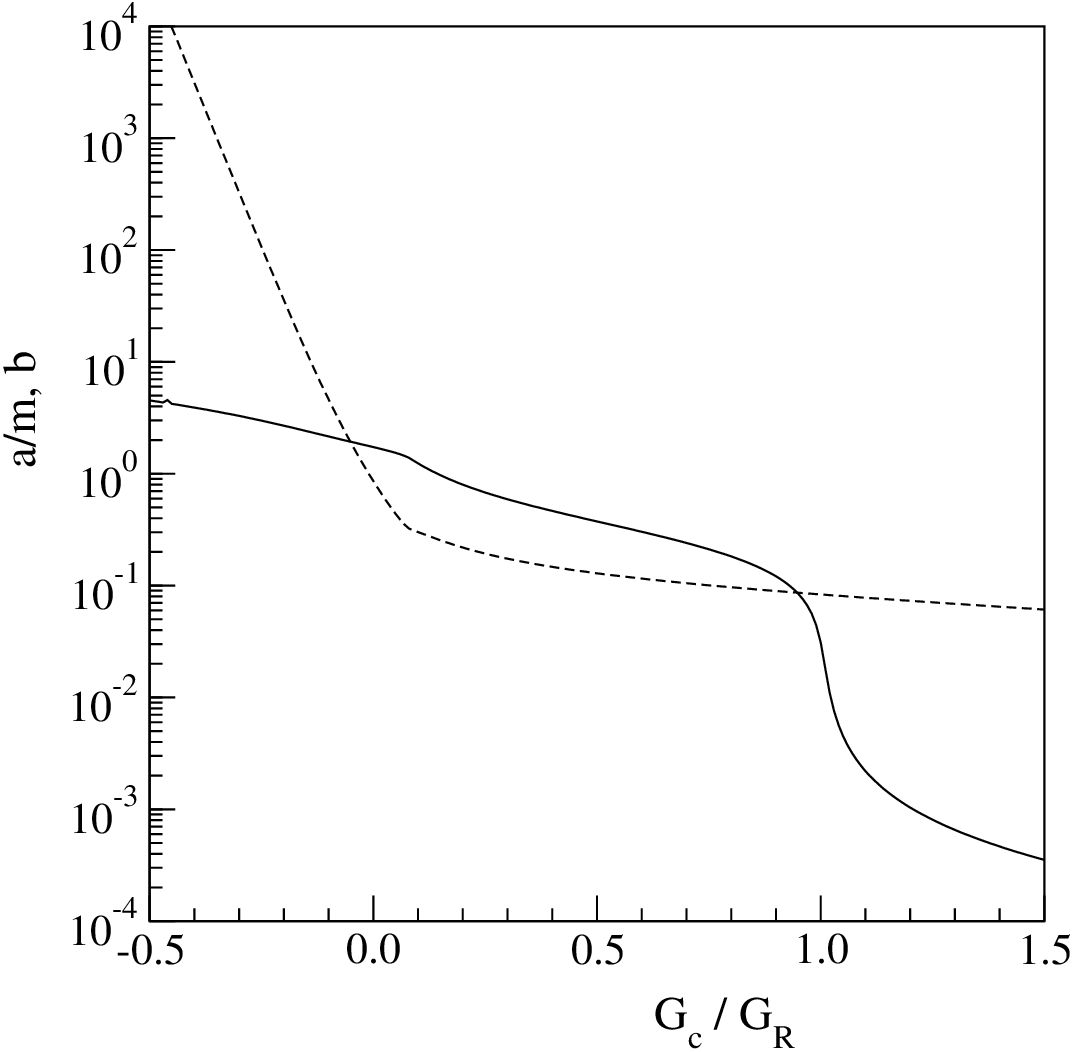}
\caption{Coefficients of the type-II NG boson dispersion relation at zero
temperature. $a/m$ is plotted by the solid line and $b$ by the dashed one.}
\label{Fig:GBdispersion}
\end{figure}

Another hint on the range of momentum where the universal predictions of the
broken symmetry apply is provided by the size of the coefficients $a,b$ defined
by Eq. \eqref{ab_coefficients}. In Fig. \ref{Fig:GBdispersion} we plot these
coefficients against the inverse coupling. Apparently, the coefficient $a$
which gives rise to the quadratic onset of the dispersion of the type-II NG
boson, is suppressed in both the BCS and RBEC regimes. In the BCS limit, this is
because the populations of the gapped and ungapped quarks are almost the same
and hence the density of the non-Abelian charge $\lambda_8$ goes to zero. In
the RBEC limit, the reason is the above-mentioned particle--antiparticle
symmetry.

\section{Conclusion}
In this paper we studied the relativistic BCS--BEC crossover in a class of
NJL-type models. Within the mean-field approximation, we investigated the
superfluid phase with particular focus on the collective bosonic excitations.
The formalism we developed is general enough to cover all possible spin-zero
pairing patterns. The calculation of the thermodynamic potential up to one
bosonic loop splits into a kinematic and an algebraic part. The kinematic part
is universal and the choice of the pairing pattern enters only through the
values of the gap parameters in the fermionic quasiparticle dispersion
relations. The algebraic part bears all necessary information about the pairing
pattern and can be done at once by means of a simple trace.

Despite the apparent generality of our model Lagrangian \eqref{Lagrangian}, two
extensions are necessary before the model can claim phenomenological relevance
for relativistic BCS--BEC crossover. First, it would be desirable to have
different chemical potentials for the individual fermion flavors in order to be
able to account for external (e.g. magnetic) fields, or for the requirement of
overall charge neutrality. Second, we should also include fermion--antifermion
pairing channels so that we can describe the constituent quark masses and the
competition between the chiral and diquark condensates.

As a particular application, we described the BCS--BEC crossover in the
two-flavor color superconductor. We investigated in detail the spectrum of NG
bosons. Their low-momentum behavior was checked to comply with the general
consequences of spontaneous symmetry breaking at finite charge density.
Concretely, the five broken-symmetry generators correspond to one NG boson with
a linear dispersion relation, and a doublet of NG bosons with a quadratic
dispersion relation. This agrees with the general counting rules for the NG
bosons in Lorentz-noninvariant systems
\cite{Nielsen:1976hm,Brauner:2005di,Brauner:2007uw}.

At the very end it should be emphasized once again that the mean-field
approximation itself is not sufficient to describe the crossover accurately.
However, the investigation of the spectrum of bosonic collective states that we
performed here, is to be understood as a necessary first step towards its
systematic improvement.

\begin{acknowledgments}
The author is indebted to H. Abuki, J. O. Andersen, J. Ho\v{s}ek, D. H. Rischke, and
A. Sedrakian for fruitful discussions and/or critical reading of the manuscript. He is
also grateful to P. Koloren\v{c} and D. Parganlija for technical help on computing
issues. The present work was in part supported by a Research Fellowship from
the Alexander von Humboldt Foundation, and by the GA CR grant No. 202/06/0734.
\end{acknowledgments}

\appendix
\section{Some useful formulas}
\subsection{Energy projectors}
\label{App:projectors}
We use the standard energy projectors $\Lambda_{\vec k}^e$ \cite{Rischke:2004rw},
associated to the
solutions of the Dirac equation for a free fermion of mass $m$, with energy
$e\eps_{\vec k}$, $e=\pm$,
$$
\Lambda_{\vec k}^e=\frac12\left[1+\frac e{\eps_{\vec k}}\gamma_0(\vec\gamma\cdot\vec
k+m)\right].
$$
We use the following properties of the projectors,
\begin{gather*}
\slashed k\pm\mu\gamma_0-m=\gamma_0\sum_{e=\pm}(k_0\pm\mu-e\eps_{\vec
k})\Lambda_{\vec k}^e,\\
\gamma_0\Lambda_{\vec k}^e\gamma_0=\Lambda_{-\vec k}^e,\quad
\gamma_5\Lambda_{\vec k}^e\gamma_5=\Lambda_{-\vec k}^{-e}.
\end{gather*}

\subsection{Matsubara sums}
\label{App:Matsubara}
Computation of the fermion loop contribution to the inverse boson propagator,
involves the following two types of fermionic Matsubara sums \cite{Schmitt:2006sw},
\begin{multline*}
T\sum_n\frac{i\omega_n-a}{(i\omega_n)^2-A^2}\frac{i(\omega_n-\Omega_N)+b}
{(i\omega_n-i\Omega_N)^2-B^2}\\
=-\frac1{4AB}\sum_{e_1,e_2}\frac{(A+e_1a)(B+e_2b)}
{i\Omega_N+e_1A+e_2B}\frac{f(e_1A)f(e_2B)}{n(e_1A+e_2B)},
\end{multline*}
\begin{multline*}
T\sum_n\frac{1}{(i\omega_n)^2-A^2}\frac{1}
{(i\omega_n-i\Omega_N)^2-B^2}\\
=\frac1{4AB}\sum_{e_1,e_2}\frac{e_1e_2}
{i\Omega_N+e_1A+e_2B}\frac{f(e_1A)f(e_2B)}{n(e_1A+e_2B)},
\end{multline*}
where $f(x)=1/(e^{\beta x}+1)$ and $n(x)=1/(e^{\beta x}-1)$ are the Fermi and
Bose distributions, respectively, and $\Omega_N$ is an arbitrary (external) bosonic
Matsubara frequency.

In the main text, we use these two respective formulas with $a=e\xi_{\vec k}^e$,
$b=f\xi_{\vec q}^f$, $A=E_{\vec kr}^e$, and $B=E_{\vec qs}^f$ to define the
quantities $I^{ef}_{rs}(\vec k,\vec q;i\Omega_N)$ and $J^{ef}_{rs}(\vec k,\vec
q;i\Omega_N)$. Let us note that they possess the following symmetry,
\begin{align*}
I^{ef}_{rs}(\vec k,\vec q;i\Omega_N)&=I^{fe}_{sr}(\vec q,\vec k;i\Omega_N),\\
J^{ef}_{rs}(\vec k,\vec q;i\Omega_N)&=J^{fe}_{sr}(\vec q,\vec k;i\Omega_N).
\end{align*}

\end{document}